# 28 Eshel Ben-Jacob: A Unique Individual in the Science of Collective Phenomena

Herbert Levine

It is my honor to write this chapter remembering the many contributions of Eshel Ben-Jacob to the field of collective behavior. I will take a personal perspective in this endeavor. That is, rather than summarize at length all the various contributions Eshel made to our science, I will focus on my understanding of his unique approach to this field obtained through my many encounters with him over the past decades.

I have broken the essay down into chronological periods. The time line begins in the early 1980s, when I first met Eshel by way of initial competition and then collaboration on pattern formation in the nonliving world. The trajectory then proceeds to our attempts to find the right biology problem, culminating in Eshel's bacterial colony phase. Lessons learned from the bacteria then became tools with which to attack other living systems, including eventually a rather bold attempt to make sense of the collective behavior of cancer cells as they progress toward metastasis. Throughout, I hope to do justice to the uniqueness of Eshel's research style and to his unyielding dedication to showing how to "let the complex be simple."

## Pattern Formation as an Entry to Complexity

Eshel loved the complexity of nature, whether he was staring at bacterial colonies or at the natural rock formations near Petra, whether studying the surface-tension-induced churning of immiscible liquids or the evolution-caused churning of network computations in cancer cells. Thus, it was only natural that he was drawn to the field of self-organization, the ability of physical systems out of equilibrium to create complex spatiotemporal patterns.

I first met Eshel when we both entered this field in the context of the study of dendritic crystal growth of the type made popular by the snowflake. He was a postdoctoral fellow at the Institute for Theoretical Physics (ITP) at the University of California, Santa Barbara, and I had just taken up a research position at an industrial lab. We and our collaborators were all enticed by a review article recently written by James Langer (1980), Eshel's postdoctoral supervisor, on a theoretical approach for understanding why dendrites form and why they grow at certain rates with a limited set of variations. Independently, both our groups hit upon the same idea; we endeavored to construct a simplified mathematical model which would enable us to cut through what at the time was the intractability of solving the full equations of the system. It would be the first time but certainly not the last that our paths converged.

After two years of hard work and back-and-forth debate over whose approach was better, an amazing set of insights emerged. Actually, even the emergence in itself was amazing as both of our groups were invited to speak at a meeting at UC Santa Barbara. Neither of us wanted to reveal in advance that we had solved the problem, leading hopefully to an untypically dramatic talk. As each group gave their talk, it became clear that we had both come up with the same solution (Ben-Jacob et al. 1984; Kessler et al. 1985)! Langer had been right about the questions but wrong about the answers. Collectively, we had developed a new theory of snowflake growth.

This is not the place to go into the mathematics behind our discoveries, which I named the principle of "microscopic solvability" (Kessler et al. 1988). Instead, I want to focus on one of the consequences of our ideas, as it forms the backdrop of the next part of the story. Our theory pointed out that the ordered crystal structures could only form if there was an anisotropy to the growth laws, such anisotropy as occurs naturally for any crystalline material, which therefore grow dendritically. What would happen without such a symmetry breaking?

In parallel, others had been investigating disordered growth patterns, the most exciting of which was the fractal structure resulting from the simple process of diffusion-limited aggregation (DLA) (Witten and Sander 1981). Our work unified this subfield with that of dendrite growth—in the absence of anisotropy, the same growth dynamics would make fractals. Conversely, imposing anisotropy artificially on a naturally symmetrical system would create dendrites. The macroscopic pattern could be controlled by the right type of small perturbations on the microscopic scale.

But how could one test this? The answer involved our next convergence. Eshel came to visit my company's lab as we started to collaborate after the ITP meeting. One of the disordered fractal-generating systems then on our minds was that of multiphase fluid flow, where water tries to push more viscous oil, a problem of immediate interest to my oil industry employer. A standard laboratory realization of the system places fluid between two narrow glass plates; as air is pushed through a hole in one of the plates, it displaces the fluid and makes a fractal pattern. During the course of Eshel's visit, I mentioned to him that one of our researchers had created a version of this experiment where the flow would have to take place through a channel network; he did this by creating a grooved plate and attaching a flat plate right on top of it, leaving only channels. Eshel immediately realized that this grooved plate could be used differently, leaving a traversable gap and hence having the grooves only slightly bias the flow—grooves were our artificial anisotropy! We ran to the lab and within 15 minutes had found the first-ever dendritic pattern in multiphase

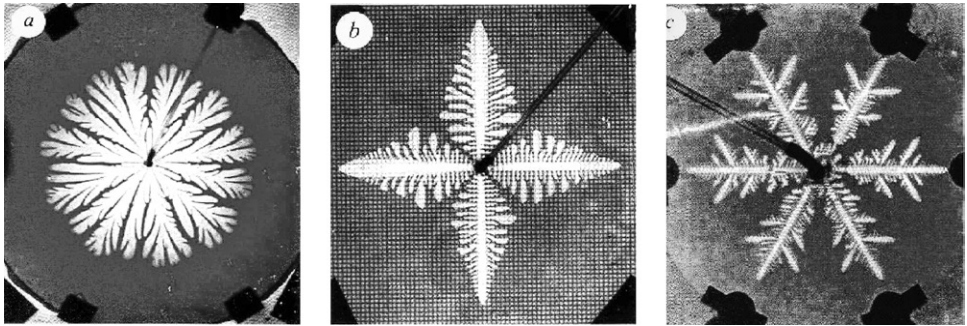

**Figure 28.1**
Patterns formed from the injection of air into a viscous fluid. (a) Isotropic pattern. (b) Imposing four-fold symmetry by a grooved plate creates a dendritic structure. (c) Adding six-fold anisotropy allows for the creation of a snowflake crystal mimic. Pictures from the Ben-Jacob lab at Tel Aviv University.

flow (see figure 28.1). This led to our first paper together in 1985 (Ben-Jacob et al. 1985).

Whenever scientists develop a new conceptual framework, and a new set of tools to go with it, a number of previously difficult systems all of a sudden become possible to analyze. After 1985, Eshel and I both entered this phase of our pattern-formation research. Building upon the aforementioned impromptu experiment, Eshel started a pattern-formation lab in his new location at the University of Michigan to explore a range of systems (Ben-Jacob and Garik 1990). We helped develop some new computational tools based on our new framework (Collins and Levine 1985; Kupferman et al. 1992). This phase of research is good for the ego, as one's papers are accepted, grants funded, and travel desires met by exciting invitations. Ultimately, however, it becomes time to get on to the next major challenge. It was clear to us that this next challenge had to involve self-organization in the living world.

## In Love with Bacteria

As anyone interested in self-organization will immediately admit, the range of such structures formed by physical and chemical systems pales in comparison with what occurs in the world of the living. The role of nonequilibrium physics in biological self-organization is masked by an increasingly unwieldy mass of biological detail. The beautiful snowflake pattern is obviously not encoded in each water molecule and hence clearly results from their collective behavior. But to what extent is the same true for biological patterns such as zebra stripes or plant leaves? Cells are much more capable than molecules, and in any case molecules do not evolve and these patterns are not functional. How could one cross this gulf to begin the study of living structures?

Eshel and I discussed this issue at length on a long drive back to my home in Connecticut from a Gordon Conference. Eshel's thoughts during that period are captured in a cartoon drawing from that era (see figure 28.2). We resolved to look for biological realizations of the type of physical patterns we had come to understand. Our thinking was that starting with a case where biological entities behaved like "dumb" molecules and recapitulated physical patterns would serve as a bridge to cases where the innate intelligence of the constituents was playing an equal role in their collective coordination. A few years later, Eshel had found his bridge. A paper by a Japanese group (Fujikawa and Matsushita 1989) showed that bacterial colonies grown under stress could make fractal patterns. Under the right conditions the cells could indeed act like the dumb particles of the DLA model. Eshel, now at Tel Aviv, started a microbiology effort to grow the bacteria.

At first, the bacterial project looked just like nonequilibrium physics. Eshel created a mathematical model to explain why the strain of bacteria he was using would create growth patterns and published the result in *Nature* (Ben-Jacob et al. 1994). Eshel and others measured fractal dimensions, growth rates as a function of nutrient, and other sensible physical measures. However, he wanted more, wanted to create a system where information on the individual scale could play a more significant role, a role more worthy of a living entity. Remarkably, as he varied the stresses imposed on the bacteria, they began to show their stuff. All manner of patterns burst forth on the Petri dishes (Ben-Jacob et al. 2000); these patterns eventually graced the cover of *Scientific American* (Ben-Jacob and Levine 1998) and the documentaries of the BBC (2014). An example of these remarkable patterns is shown in figure 28.3. Many questions immediately presented themselves. First, why had no previous researcher seen this vast menagerie? How could bacteria evade what had been thought of as the ironclad rule of physics that two-dimensional systems cannot break continuous symmetry—the self-organized motion of the "vortex" phase appeared to do just that. Why did some patterns show remarkable chirality? What modes of communication could account for the much more coordinated actions of the cells, beyond what would be expected for short-range physical interactions of the molecular type? The central phase of Eshel's career was born!

The answers turned out to be as interesting as the questions. No one had seen these patterns before because Eshel had found a new species of bacteria. He promptly named his initial strain *Paenibacillus dendritiformis* (Tcherpakov et al. 1999) after its most obvious property. He was incredibly proud of having, as he put it, the first physics lab that had ever isolated a new bacterium. As far as the no-symmetry-breaking violation, it turned out that this was true only for equilibrium systems; living systems, and even other systems that are being actively driven by external sources of energy, need not obey. Eshel's paper on the subject (Vicsek et al. 1995), written together with Tamas Vicsek and others, launched the new field of active matter and has been cited by thousands. What about coordination? Eshel speculated that the cells exchanged a variety of chemical signals, which allowed each individual to interpret the actions of the collective vis-à-vis its own state. These types of information-passing channels were well-known in the world of advanced multicellular organisms but were assumed absent in bacterial systems where every cell was presumed to act on its own. Eshel argued that this was thought to be true only because microbiologists had long grown accustomed to laboratory strains grown under nonstressful conditions; the bacteria would reveal their secrets only under duress. Some prescient biologists such as James Shapiro had argued for coordination (Shapiro 1998), but it was an idea far from mainstream. A few years later, the mainstream caught up with the discovery of quorum sensing factors and eventually also other such signals (Waters and Bassler 2005).

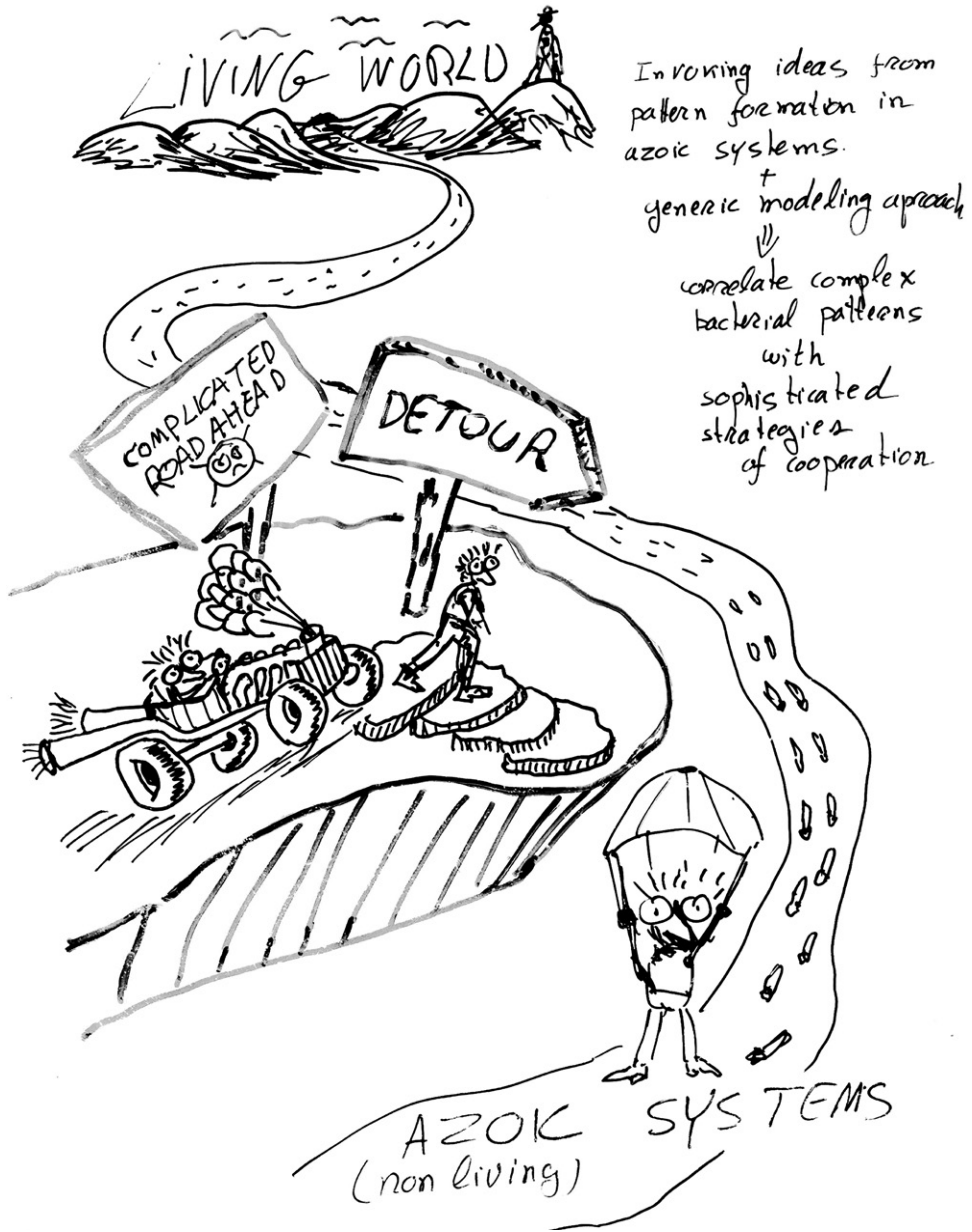

**Figure 28.2**
The road to biology, unpublished cartoon, by Eshel Ben-Jacob.

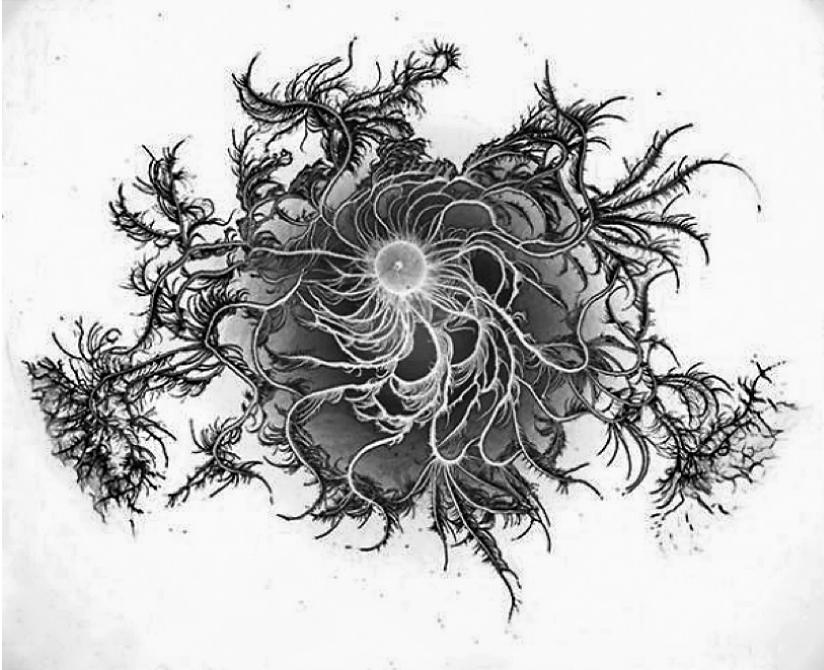

**Figure 28.3**
*Paenibacillus dendritiformis* pattern. From the first instant he saw this type of colony, Ben-Jacob called it bacteria art; "Without knowing anything, you'll feel the sense that there is drama going on." Ben-Jacob lab at Tel Aviv University, unpublished.

What about the chiral structure? In nonliving systems, we had jointly shown that microscopic changes could affect macroscopic patterns. We realized that this could enable a unique role for the cell and its gene-based information. By changing key parameters, the cell would help alter the whole pattern. For the chiral structure, cells divided only after growing much longer, increasing the aspect ratio by a large factor. Interacting long rods had the capability of enhancing small left–right symmetry breaking effects much like interacting crystal molecules had the capability of enhancing rotational anisotropy effects. Microscopic solvability became, in Eshel's language, a macro–micro interplay (Ben-Jacob and Levine 2006) where cells could act to create functional structures on the whole-colony scale. Exactly how much of this is true, whether the myriad of patterns are somehow a functional response to sensed environmental conditions or merely a result of the physics of self-organization, remains unresolved for the rest of us. However, Eshel was completely convinced and argued that the bacteria were exhibiting rudimentary intelligent behavior (Ben-Jacob et al. 2004). Even as we argued, I helped Eshel formulate some of these provocative ideas as a way of spurring the field forward. The road to the living world was now open.

**A Bacteria-Centered View of the Brain**

As the bacteria work continued, Eshel began to sense that the ideas he had developed could have broader implications. Just as the bacteria had been underestimated, in particular with regard to their use of information to help fashion their mutual environment, so too other parts of biological systems had been underestimated. Just as the overall structure of the colony made sense when viewed through the lens of the micro–macro interplay, so too would other even more complex biological systems make sense when one allowed the individual actor to take some responsibility for the behavior of the collective. With this perspective, Eshel turned to the brain.

For the brain, the glial cells would play the role of the bacteria. These cells had been relegated in the approach of most of the neurobiology community to a supporting role; they provide food for the neurons and clean up excess chemicals in the neighborhood of the synapses to allow for crisp transmission of electrical signals. But Eshel was not convinced. He felt that the responses of these cells to synaptic activity, mediated by the nonlinear dynamics of calcium signals, were perfectly attuned to what would be needed for the establishment of the type of communication channel he had argued for in the bacterial case. These cells could coordinate nerve firing on a timescale relevant for memory, for example, and therefore must be doing exactly that in the brain.

To help demonstrate his ideas, Eshel again remade himself, this time into an experimental neurobiologist focused on cultured networks (Segev et al. 2004). He showed how these neural systems, neurons plus glial cells, led to dynamical patters of activity; space was augmented by time as a canvas for the self-organization. One high point was his demonstration of how patterns could be imprinted into the neural network by proper chemical stimulation (see figure 28.4); this work (Baruchi and Ben-Jacob 2007) received a great deal of press coverage as it tapped into the popular idea that useful artificial neural systems may one day be developed. Modeling helped show that the patterns required the glial coordination (Volman et al. 2007). I helped flesh out these ideas because of another

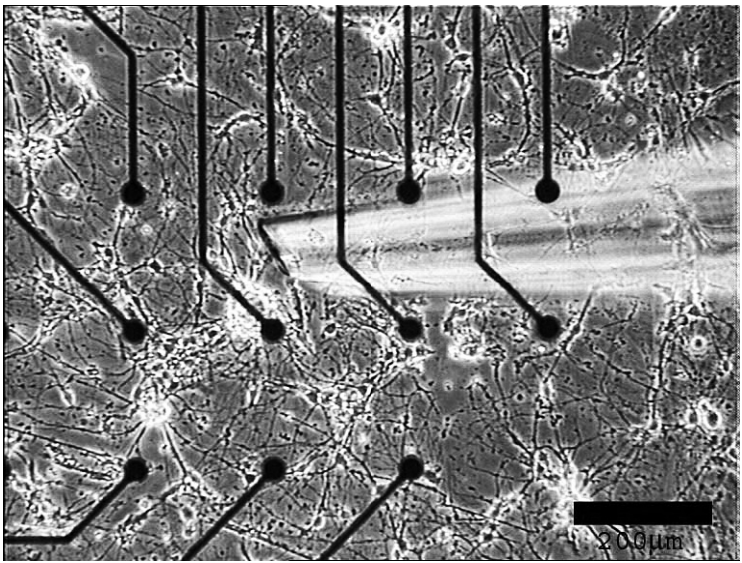

**Figure 28.4**
Stimulation of neurons leads to imprinted activity patterns; Ben-Jacob lab at Tel Aviv University.

convergence—I had been working on the role of calcium dynamics in other cell-biology contexts (Falcke et al. 2000).

Looking back, I think it is fair to say that Eshel made much less impact on the study of neural systems than on the science of bacterial colonies. Many physicists took to studying bacteria and, while not necessarily agreeing with all of Eshel's perspectives, appreciated the groundbreaking ideas he had inserted into the field and his push toward a collective rather than an individualistic approach to the colony. Neurobiology is a much harder field for physicists to just pop into, and consequently provides a less receptive audience. And, the ideas being proposed here were, in the end, less revolutionary. Neurons already performed their magic via collective effects, and adding glial interactions seemed like a quantitative rather than a qualitative insight. Cultured networks are not universally accepted as useful experimental realizations of relevant neuronal behavior, and the experiments on induced patterns did not scale up to more complex tasks.

Perhaps more time was needed to really alter mainstream thinking in neuroscience, a vast enterprise that always seems on the verge of major breakthroughs. However, time was not to be had. Eshel's health, a major issue since his days at Michigan, took a decided turn for the worse. Eshel had cancer and so began the last phase of his career, his personal and professional battle against the "emperor of all maladies" (Mukherjee 2011).

**Eshel versus Cancer, a War on Several Battlegrounds**

Eshel once told me that he felt that his entire previous career had been a training ground to prepare him to understand and eventually defeat cancer. Surveying the landscape of cancer research, he found a familiar picture—cancer cells were thought to be solitary individual actors whose mutations allow them to ignore their proper role in the multicellular organism and hence just grow. Very few researchers spoke of cell–cell cellular cooperation, environmental shaping by sensing and acting, collective motility—it was a return to the role supposed to have been played by the nonstressed resource-abundant bacterial cell. It appeared inconceivable that this crippled loner cell could somehow evade the best efforts of the biomedical community, which had dedicated decades to designing sophisticated cancer therapies.

It was our last convergence. I too had done some thinking about the nascent physics of the cancer field. It was time, in my opinion, to extend the reach of nonequilibrium physics to issues of biomedical, not merely biological, significance. This had happened in the field of cardiac disease, thanks initially to the work of individuals such as Arthur Winfree (2001), and I was impressed at the progress being made. Cancer research, by contrast, seemed weighted down by the sheer mass of collected facts and in need of a dose of physics-style focus on the essentials. Armed with our biases, we set out on what would turn out to be our last joint scientific journey.

Our first task was to explain, mostly to ourselves, how our thoughts about cancer had emerged from our experience with bacteria. We wrote an opinion article for *Trends in Microbiology* (Ben-Jacob et al. 2012), outlining how cancer research needed to focus more on the intelligence of the cancer cell, its coordination through signaling with other cancer cells and noncancer cells in its vicinity, and how the paradigm of killing the soldiers needed to be updated to disrupting their command and control, launching a cyber war against a smart and determined foe (see figure 28.5). For us, writing this paper was a trip back in time to the days when we used to argue about nonequilibrium physics and bacterial intelligence. We have since progressed to more concrete efforts in the cancer field, focusing on cellular decision-making and on collective motility as a path toward metastatic spread (Jolly et al. 2015). The origins of these ideas in our studies of microorganisms remain clear to me but perhaps are more downplayed now than in the beginning.

While our professional efforts on cancer biology intensified, Eshel's private battle against the disease itself raged on. As was expected by all who knew him, Eshel would chart his own course here as well. Based on ideas regarding cancer metabolism, Eshel became enamored with hyperbaric oxygen treatment and with the diabetes drug metformin. Based on collaborations with oncologists that were to his liking (the more adventurous ones, of course), he made up his own drug regimen and resisted surgery. Eshel tried to have his history published as a case report documenting the efficacy of his version of personalized medicine, but the oncology community has not yet been persuaded.

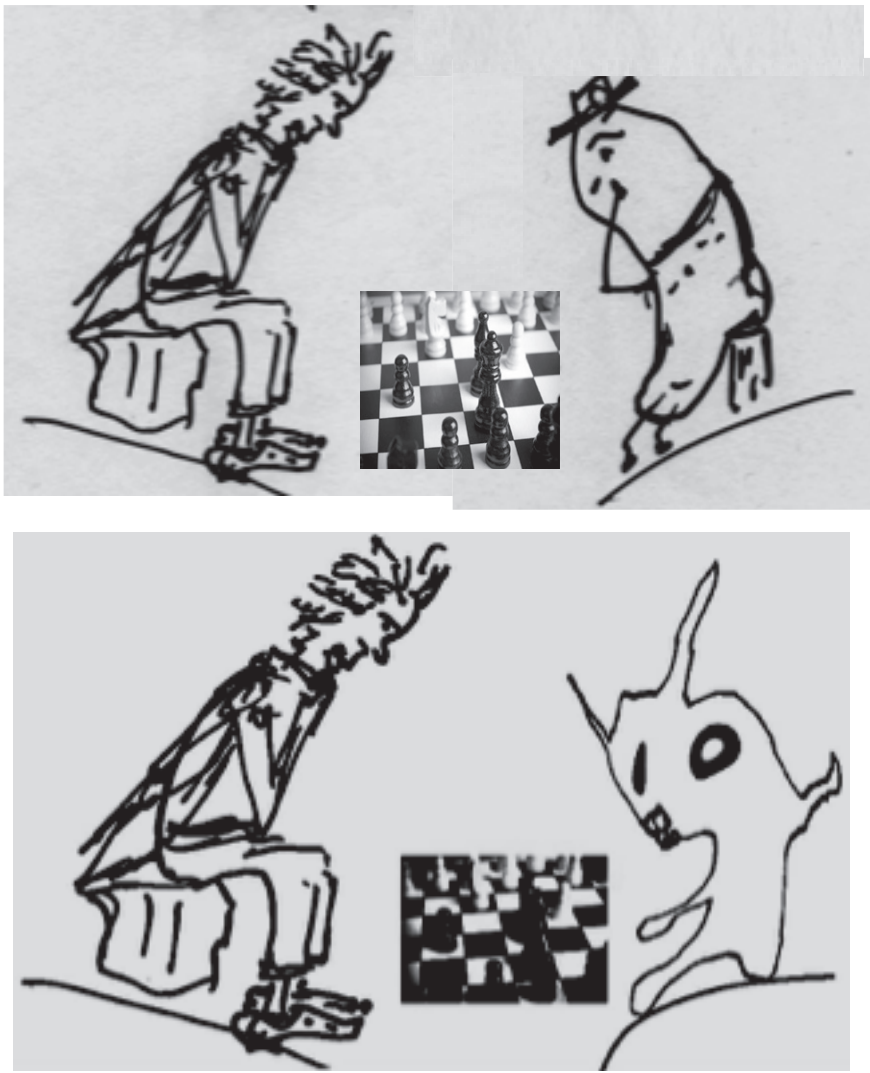

**Figure 28.5**
"By thy wise counsel shalt thou make war" (Proverbs 24:6); cyber war against pathogenic bacteria and against cancer. Unpublished cartoons by Eshel Ben-Jacob.

Throughout, Eshel never stopped thinking, working, and advising. I have no idea as to how he had the energy to work with all his collaborators, on all his different interests, at all hours of the day or night. Toward the end, he acted as if he knew he was running out of time and needed to make just one more connection before all the pieces of the cancer puzzle would fit neatly together. By now, any reader without firsthand knowledge is surely assuming that cancer gained and kept the upper hand and Eshel slowly slipped away. Actually, he seemed to be winning the battle when he died suddenly of cardiac failure. In an ironic twist, cardiac dynamics was one of the few areas of biological physics that Eshel did not work on.

**Epilogue**

In June 2015, I sat in a room at Tel Aviv University filled with researchers from the fields of microbiology, neurosciences, linguistics, economics, immunology, cancer biology, and physics. Many knew only a few others in the room, but all were connected through Eshel. In his path through science, Eshel had picked up many fellow travelers who saw him as someone who might breathe new life into their own subject. It was only a few days after his passing, and everyone was still in shock. The discussion inevitably turned to what could be done to keep the group together. I was pessimistic. Only Eshel could have created such a collection of people and scientific investigations, and Eshel was now gone. The group would inevitably fragment, for such appears to be the most stable dynamical process of academic science. It was only Eshel's boundless energy and enthusiasm that turned the tide in favor of the broadest possible collective. And, as we all go our separate ways, let us remember him fondly.

**Acknowledgments**

I acknowledge the countless hours over the last three plus decades that I have spent discussing science with Eshel. This work was supported by National Science Foundation (NSF) Center for Theoretical Biological Physics (NSF PHY-1427654) and CPRIT (Cancer Prevention and Research Institute of Texas) Scholar in Cancer Research of the State of Texas at Rice University.